\documentclass[preprint,aps,showpacs]{revtex4}
\usepackage{amsfonts}
\usepackage{amssymb}
\usepackage{dcolumn}
\usepackage{bm}
\usepackage{array}
\usepackage{graphicx}
\usepackage{caption2}
\usepackage{amsmath}
\usepackage{subfigure}
\usepackage{color}
\begin{document}
\title{Entangling two atoms in spatially separated cavities through both photon emission and absorption processes}
\date{\today}
\author{Peng Peng}
\author{Fu-li Li}
\email[Email: ]{flli@mail.xjtu.edu.cn}
\affiliation{Department of
Applied Physics, Xi'an Jiaotong University, Xi'an 710049, China}
\begin{abstract}
We consider a system consisting of a $\Lambda$-type atom and a
V-type atom, which are individually trapped in two spatially
separated cavities that are connected by an optical fibre. We show
that an extremely entangled state of the two atoms can be
deterministically generated through both photon emission of the
$\Lambda$-type atom and photon absorption of the V-type atom in an
ideal situation. The influence of various decoherence processes
such as spontaneous emission and photon loss on the fidelity of
the entangled state is also investigated. We find that the effect
of photon leakage out of the fibre on the fidelity can be greatly
diminished in some special cases. As regards the effect of
spontaneous emission and photon loss from the cavities, we find
that the present scheme with a fidelity higher than $0.98$ may be
realized under current experiment conditions.
\end{abstract}
 \maketitle

\section{Introduction}

Entangled quantum state is one of essential key ingredients in the
implementation of quantum communication and quantum computation
\cite{Bennett1,Ekert,Bennett2,Nielsen}. Among kinds of schemes
proposed to generate entangled states
\cite{Reina,Pellizzari1,Chuang,Cirac,Turchette}, cavity quantum
electrodynamic systems(CQEDS), which marry atomic and photonic
quantum bits together, are paid more attentions because of both its
low decoherent rate and promising feasibility to scale-up. In recent
years, several elegant works based on CQEDS have been done. In
\cite{Lloyd}, a pair of momentum- and polarization- entangled
photons is sent to two spatially separated cavities, each of which
contains a V-type atom. After the atoms absorb the photons, the
photon entanglement is transferred to the atoms. In
\cite{Feng,Duan,Browne,Duan1}, entangled states of two
$\Lambda$-type atoms individually trapped in spatially separated
cavities are generated through the interference of the polarized
photons leaking out of the cavities on a beam splitter and
co-instantaneous single-photon detections. In these schemes, in
order to guarantee the interference effect, several conditions are
required. First, the two atoms have to be simultaneously excited or
driven. Second, spatial shapes of photon pulses leaking from the
cavities should be similar. Third, the two photon paths from the
cavities to the beam splitter must be symmetrical. In addition, the
success of generating entangled atomic states is probabilistic since
the photon-state-projection detection behind the beam splitter is
performed. We also notice that the previous schemes are based on
either photon emission or photon absorption process. In this paper,
we consider a CQED system in which a $\Lambda$-type atom and a
V-type atom are trapped individually in two spatially separated
cavities that are connected by an optical fiber. This setup is
closely related to two previous schemes. In \cite{Pellizzari2},
Pellizzari proposed a cavity-fibre-cavity system to realize reliable
transfer of a quantum state. In that scheme, two three-level
$\Lambda$ atoms that are adiabatically and simultaneously driven by
a laser beam are trapped individually in two single-mode cavities
connected via an optical fibre. In \cite{Alessio}, Serafini et al
employed the similar setup in which two-level atoms are trapped in
fibre-connected cavities to realize highly reliable swap and
entangling gates. We notice that quantum information processes
realized in the fibre-connected cavity setup depends on atomic level
configurations and laser driving ways. In the present scheme,
without using any laser driving, only the $\Lambda$-type atom is
initially to be pumped in the excited state. Instead of using either
photon absorption or photon emission as done in the previous scheme,
both the photon emission of the $\Lambda$-type atom and the photon
absorption of the V-type atom are involved. In this way, an
entangled state of the two atoms can deterministically be generated
through the atom-field interaction without the photon interference
and the photon-state projection detection.

\section{model}

\begin{figure}[htbp]
    \centering
       \includegraphics[width=9cm]{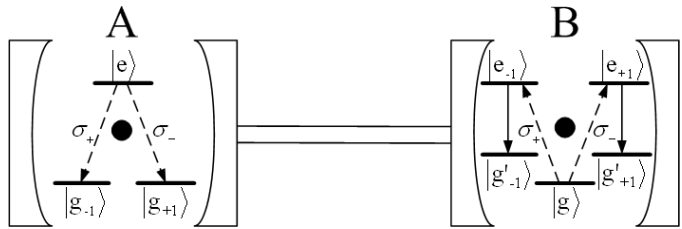}
       \caption{A $\Lambda$ type atom and a V-type atom are trapped
        in two spatially separated cavities A and B, respectively.
        The two cavities are linked through an optical fibre.}
       \label{fig:model}
\end{figure}

As shown in Fig.\ref{fig:model}, a $\Lambda$-type atom with two
degenerate states $\vert g_{-1}\rangle$ and $\vert g_{+1}\rangle$
and a V-type atom with two degenerate states $\vert e_{-1}\rangle$
and $\vert e_{+1}\rangle$ are trapped in two spatially separated
cavities that are linked by an optical fibre. The two atoms interact
with local cavity fields, respectively. In recent years, such
fiber-connected cavity systems are often empolyed for realizing
quantum information processes \cite{Pellizzari2, Alessio, Yin}.

The Hamiltonian of the whole system  can be written as
\cite{Pellizzari2}
\begin{equation}
H=H_{A}+H_{B}+H_F,
\end{equation}
where
\begin{eqnarray}
H_{A}&=&\sum_{j=\pm1}\Big[\omega^f_{A,j}a^+_{A,j}a_{A,j}+\omega^a_{A}
\sigma^z_{A,j}+\lambda_{A,j}a_{A,j}\sigma^{+}_{A,j}+\mathrm{H.c.}\Big],\\
H_{B}&=&\sum_{j=\pm1}\Big[\omega^f_{B,j}a^+_{B,j}a_{B,j}+\omega^a_{B}
\sigma^z_{B,j}+\lambda_{B,j}a_{B,j}\sigma^{+}_{B,j}+\mathrm{H.c.}\Big],\\
H_F&=&\sum_{k=1}^{\infty}\sum_{j=\pm1}\Big[\omega_{j,k} b^+_{j,k}
b_{j,k}+\nu_{j,k}b_{j,k}[a^+_{A,j}+(-1)^ke^{i\phi}a^+_{B,j}]+\mathrm{H.c.}\Big],
\end{eqnarray}
where $H_{A/B}$ is the energy of the system consisting of atom A/B
and the corresponding local cavity fields, and $H_F$ represents
the free energy of fibre modes and the interaction between cavity
and fibre modes. In (2)-(4), $a_{A/B,j}$ and $a^{+}_{A/B,j}$ are
annihilation and creation operators for photons of frequency
$\omega^f_{A/B,j}$ and polarization j(= -1,+1 corresponding to
right and left circular polarizations, respectively) in cavity
A/B, and $b_{j,k}$, $b^{+}_{j,k}$ are annihilation and creation
operators for photons of frequency $\omega_{j,k}$ and polarization
j in mode k of the fibre field, $\sigma^z_{A,j}=(\vert
e\rangle\langle e\vert-\vert g_j\rangle\langle g_j\vert)_{A}/2,
\sigma^z_{B,j}=(\vert e_j\rangle\langle e_j\vert-\vert
g\rangle\langle g\vert)_{B}/2$,  $\sigma^{+}_{A,j}=(\vert
e\rangle\langle g_j\vert)_{A}$ and $ \sigma^{+}_{B,j}=(\vert
e_j\rangle\langle g\vert)_{B}$, $\omega^a_{A/B}$ is the frequency
of the transition $\vert e/e_j\rangle\rightarrow\vert
g_j/g\rangle$, $\lambda_{A/B,j}$ is the coupling constant of atom
A/B with cavity mode $j$, and $\nu_{j,k}$ the coupling constant of
cavity mode $j$ with fibre mode $k$. Here, we assume that the
right and left circular polarization modes in both the cavities
have the same frequency $\omega$ and the interaction between the
atoms and the cavity fields is on resonance. That is,
$\omega^f_{A,\pm 1}=\omega^f_{B,\pm
1}=\omega^a_{A}=\omega^a_{B}=\omega$.

In the short fibre limit $(2L\bar{\nu})/(2\pi C) \leq 1$, where $L$
is the length of fibre and $\bar{\nu}$ is the decay rate of the
cavity fields into a continuum of fibre modes, only resonant modes b
of the fibre interacts with the cavity modes \cite{Alessio}. For
this case, the Hamiltonian $H_{F}$ may be approximated to
\begin{equation}
  \label{eq:Hf}
  H_{F} = \sum_{j=\pm1}\big[\omega b^+_{j}
  b_{j}+\nu_jb_{j}(a^+_{A,j}+a^+_{B,j})+\mathrm{H.c.}\big],
\end{equation}
where the phase $(-1)^ke^{i\phi}$ in (4) has been absorbed into the
annihilation and creation operators of the modes of the second
cavity field.

In the interaction picture, the total Hamiltonian now becomes
\begin{equation}
\label{HI1}
 H_{I}=\sum_{j=\pm 1}\big[\lambda_{A,j}a_{A,j}\sigma^{+}_{A,j}+\lambda_{B,j}a_{B,j}\sigma^{+}_{B,j}+
\nu_jb_{j}(a^+_{A,j}+a^+_{B,j})+\mathrm{H.c.}\big].
\end{equation}

\section{The generation of entangled states}

In this section, we show that an extremely entangled states of the
atoms can be deterministically generated in an ideal situation.

To explain the scheme of the generation of entangled states,
first, let us see how the entangling process works. Suppose that
at the initial time the atom A is pumped in the excited state
$\vert e\rangle_A$ and the atom B is in the ground state $\vert
g\rangle_B$. Through interacting with local cavity fields, the
atom A emits either a $\sigma_+$ or $\sigma_-$ polarized photon
and then the atom-field system is in the state
$\vert\Psi_1\rangle=\frac{1}{2}(\vert g_{-1}\rangle\vert
1_{\sigma_+}\rangle+\vert g_{+1}\rangle\vert
1_{\sigma_-}\rangle)$, where $\vert 1_{\sigma_+}\rangle/\vert
1_{\sigma_-}\rangle$ is a cavity field state with one photon in
mode $j=-1/+1$. Assume that the coupling of the cavity modes with
fibre modes is so strong that the emitted photon can be
transferred into the fibre modes before reabsorbed by the atom A.
Through the fibre, the photon enters the cavity B. After absorbing
the photon, the atom B is excited in either $|e_{-1}\rangle_B$ or
$|e_{+1}\rangle_B$. In this way, the entangled state
$|\Psi_t\rangle=\frac{1}{\sqrt{2}}(|g_{-1}\rangle_A|e_{-1}\rangle_B+|g_{+1}\rangle_A|e_{+1}\rangle_B)$
can be created. In order to protect the entangled state from
spontaneous emission, as shown in Fig.\ref{fig:model}, one may
apply a $\pi$-polarized light to the transition between $|e_{\pm
1}\rangle_B$ and $|g'_{\pm 1}\rangle_B$, and generate a stable
entangled state:
$|\psi_s\rangle=\frac{1}{\sqrt{2}}(|g_{-1}\rangle_A|g'_{-1}\rangle_B+|g_{+1}\rangle_A|g'_{+1}\rangle_B)$,
where $|g_{\pm1}\rangle_A$ and $|g'_{\pm1}\rangle_B$ are stable
states.

Now, let us go to investigate the generation of the entangled state
in detail. The time evolution of the whole system state is governed
by the Schr\"odinger equation
\begin{equation}
\label{s-equation0} i\frac{\partial}{\partial \mathrm{t}}
|{\psi_1}(t) \rangle=
  H_I |{\psi_1}(t) \rangle.
\end{equation}
Suppose that at the initial time the system is in the state
$|\psi_1(0)\rangle=|e\rangle_A|0\rangle_A|0\rangle_f|0\rangle_B|g\rangle_B$.
The state vector $|{\psi_1}(t) \rangle$ at time t can be expanded as
\begin{equation}
\label{psi1}
\begin{aligned}
|\psi_1(t)\rangle&=d_1|e\rangle_A|0\rangle_A|0\rangle_f|0\rangle_B|g\rangle_B+
d_2|g_{-1}\rangle_A|1_{-1}\rangle_A|0\rangle_f|0\rangle_B|g\rangle_B\\
&+d_3|g_{-1}\rangle_A|0\rangle_A|1_{-1}\rangle_f|0\rangle_B|g\rangle_B
+d_4|g_{-1}\rangle_A|0\rangle_A|0\rangle_f|1_{-1}\rangle_B|g\rangle_B\\&+
d_5|g_{-1}\rangle_A|0\rangle_A|0\rangle_f|0\rangle_B|e_{-1}\rangle_B+
d_6|g_{+1}\rangle_A|1_{+1}\rangle_A|0\rangle_f|0\rangle_B|g\rangle_B\\
&+d_7|g_{+1}\rangle_A|0\rangle_A|1_{+1}\rangle_f|0\rangle_B|g\rangle_B+
d_8|g_{+1}\rangle_A|0\rangle_A|0\rangle_f|1_{+1}\rangle_B|g\rangle_B\\&+
d_9|g_{+1}\rangle_A|0\rangle_A|0\rangle_f|0\rangle_B|e_{+1}\rangle_B,
\end{aligned}
\end{equation}
where $|1_{\pm1}\rangle$ represents either a photon number state
with one right(-1) or left(+1) circular polarized photon. Upon
substituting (\ref{psi1}) in (7) with the conditions
$\lambda_{A,-1}=\lambda_{A,1}=\lambda,
\lambda_{B,-1}=\lambda_{B,1}=\sqrt{2}\lambda$, and
$\nu_{+1}=\nu_{-1}=\nu$, one has
\begin{eqnarray}
&&d_1=\frac{1}{2}\Big[cos(\sqrt2\lambda
t)+1\Big]+\frac{\lambda^2}{2(\lambda^2+\nu^2)}\Big[cos(\sqrt{1+\frac{\nu^2}{\lambda^2}}\sqrt{2}\lambda t)-1\Big],\\
&&d_2=d_6=-\frac{i}{2\sqrt2}\Big[sin(\sqrt2\lambda
t)+\frac{\lambda}{\sqrt{\lambda^2+\nu^2}}sin(\sqrt{1+\frac{\nu^2}{\lambda^2}}\sqrt{2}\lambda t)\Big],\\
&&d_3=d_7=\frac{\lambda\nu}{2(\lambda^2+\nu^2)}\Big[cos(\sqrt{1+\frac{\nu^2}{\lambda^2}}\sqrt{2}\lambda t)-1\Big],\\
&&d_4=d_8=\frac{i}{2\sqrt2}\Big[sin(\sqrt2\lambda
t)-\frac{\lambda}{\sqrt{\lambda^2+\nu^2}}sin(\sqrt{1+\frac{\nu^2}{\lambda^2}}\sqrt{2}\lambda t)\Big],\\
&&d_5=d_9=\frac{1}{2\sqrt2(\lambda^2+\nu^2)}\Big[\nu^2+\lambda^2cos(\sqrt{1+\frac{\nu^2}{\lambda^2}}\sqrt{2}\lambda
t)-(\lambda^2+\nu^2)cos(\sqrt2\lambda t)\Big].
\end{eqnarray}
We notice that $d_1=d_2=d_3=d_4=0$, $d_5=d_9=1/\sqrt{2}$, and the
system is in the state
$|\psi_1\rangle=\frac{1}{\sqrt{2}}(|g_{-1}\rangle_A|e_{-1}\rangle_A+
|g_{+1}\rangle_A|e_{+1}\rangle_A)|0\rangle_A|0\rangle_f|0\rangle_B$
when $\sqrt2\lambda t=m\pi (m=1,3,5,...)$ and
$\sqrt{1+\frac{\nu^2}{\lambda^2}}=n (n=2,4,6,...)$. In this state,
the atoms are completely separated from the cavity fields and in
the extremely entangled state $\vert\Psi_t\rangle$.

\begin{figure}[htbp]
 \centering
\mbox{\subfigure[]{\includegraphics[width=7.5cm]{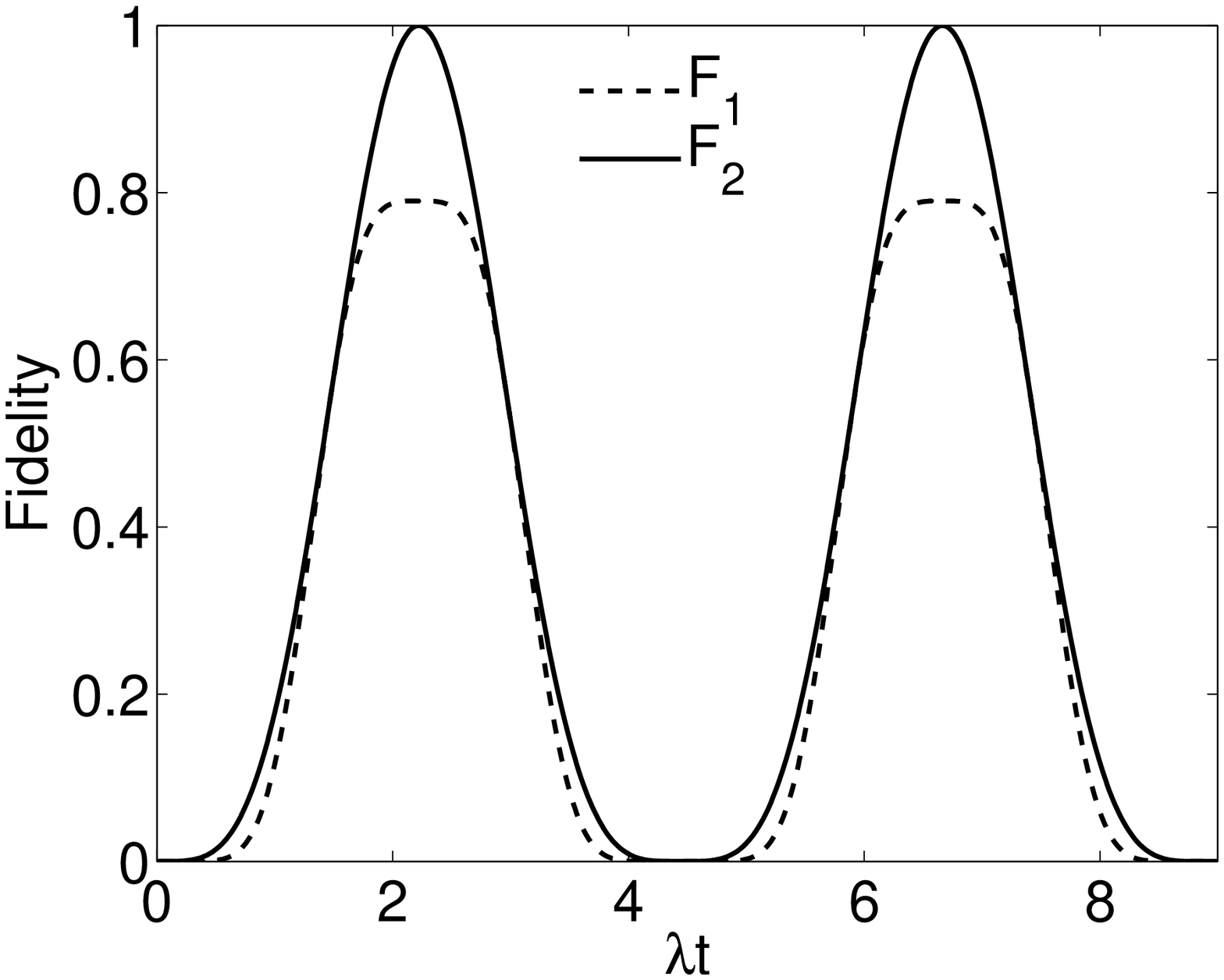}}
\subfigure[]{\includegraphics[width=7.5cm]{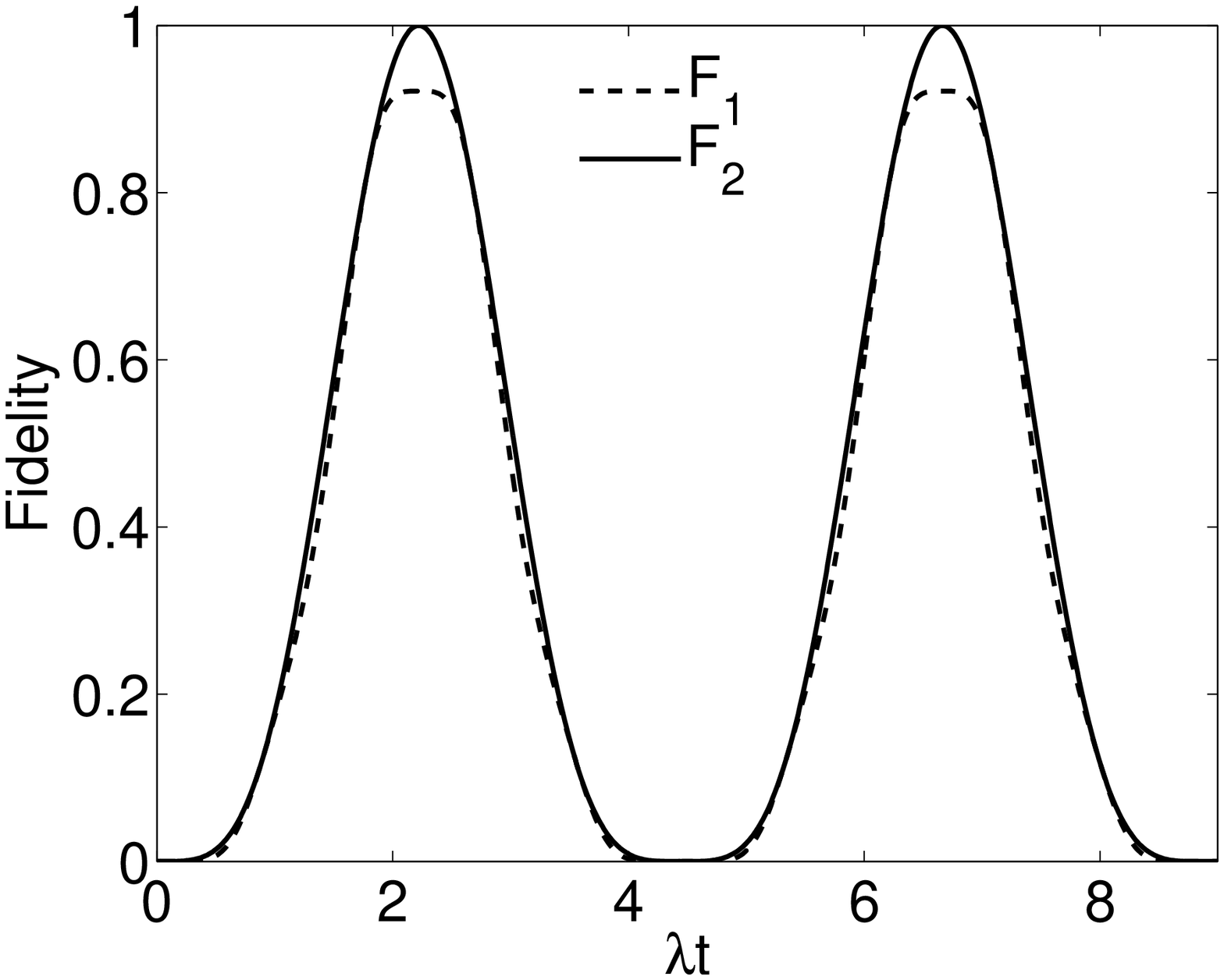}}}\\
\mbox{\subfigure[]{\includegraphics[width=7.5cm]{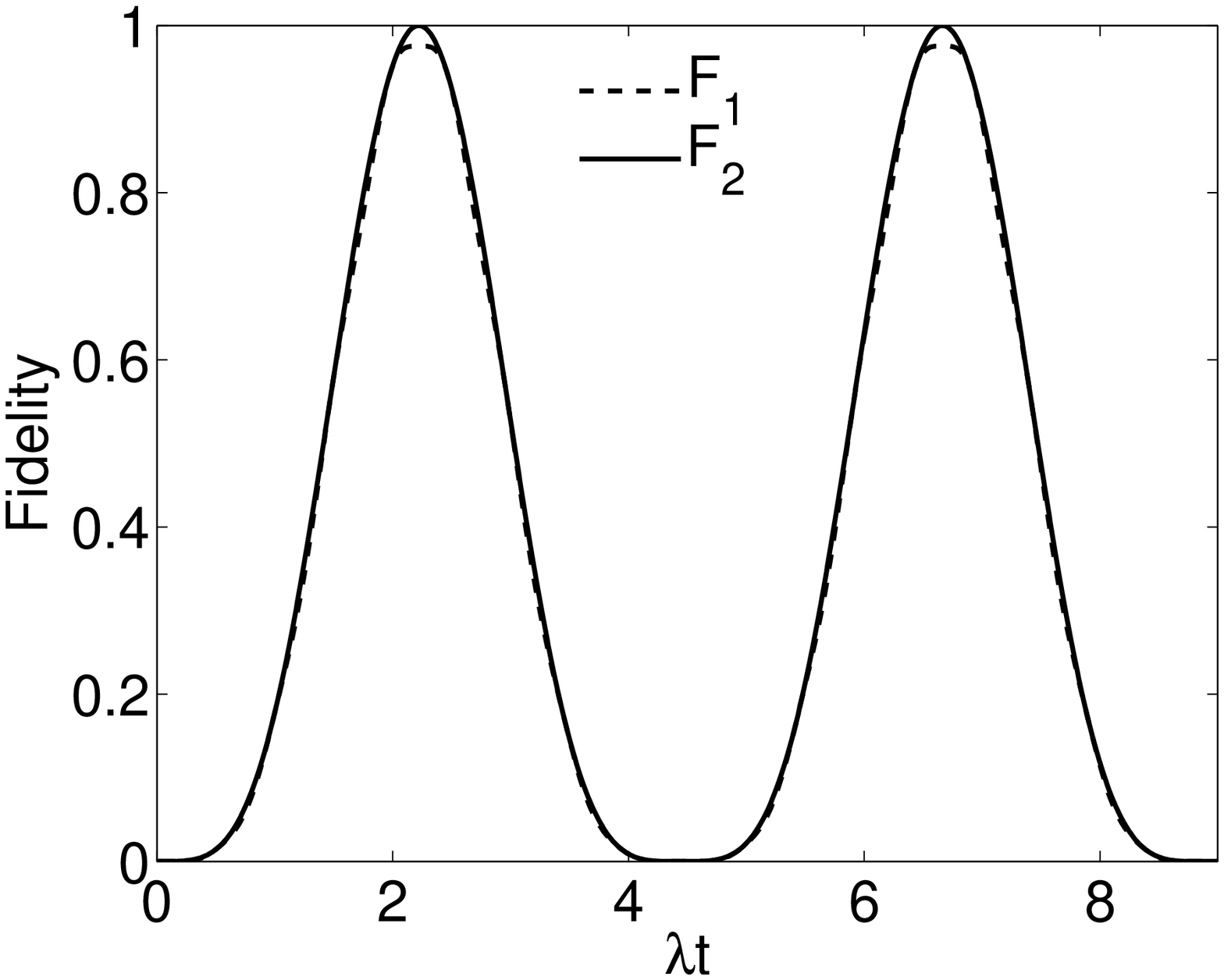}}
\subfigure[]{\includegraphics[width=7.5cm]{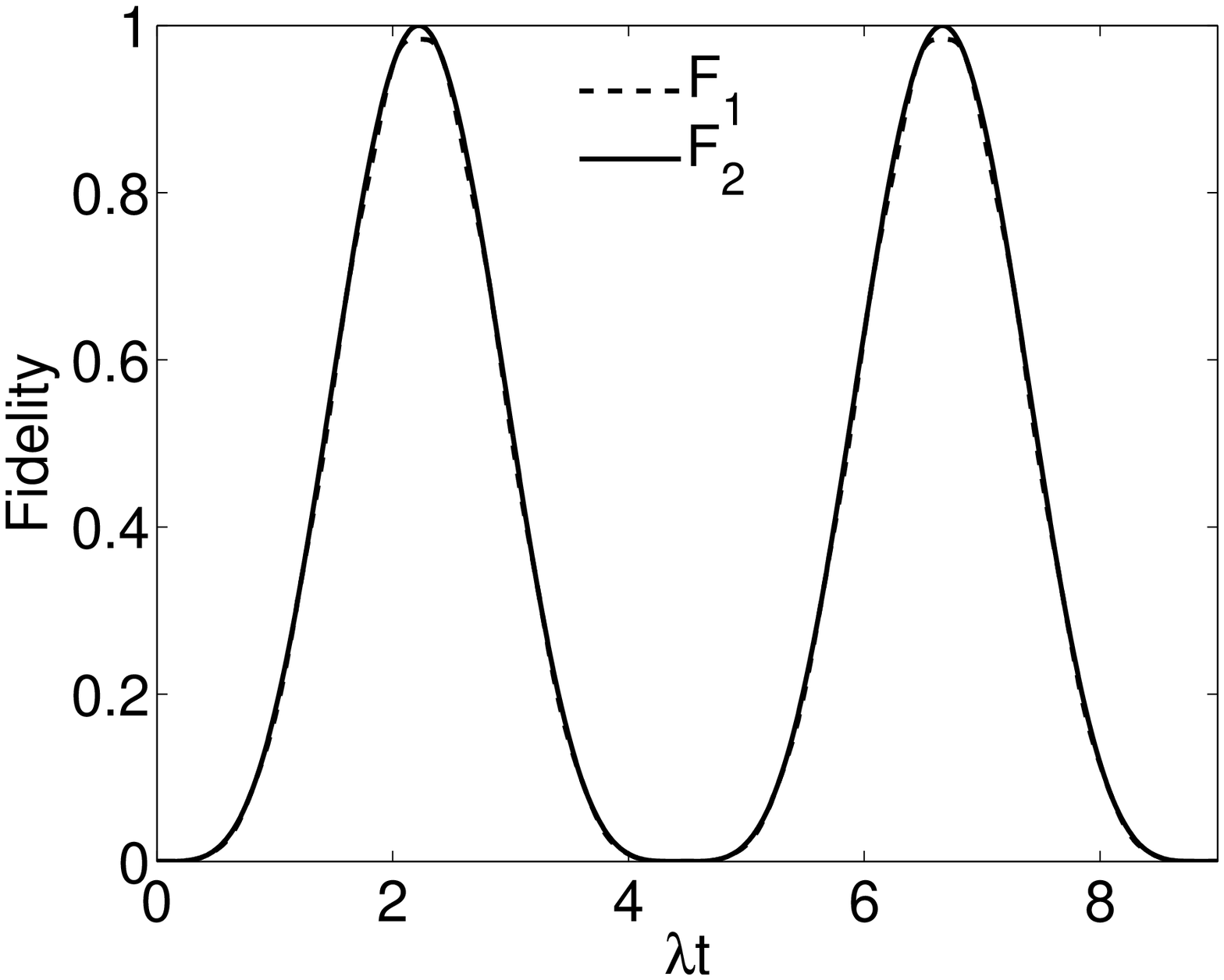}}}\\
       \caption{(a)$\nu=\sqrt{8}\lambda$;(b)$\nu=\sqrt{24}\lambda$;
       (c)$\nu=\sqrt{80}\lambda$;(d)$\nu=\sqrt{120}\lambda$}
       \label{nodis}
\end{figure}

In Eqs. (9)-(13), one may also find that if $\nu \gg \lambda$ the
conditions $d_1=d_2=d_3=d_4=0$ and $d_5=d_9=1/\sqrt{2}$ can also
be satisfied at the time $\sqrt2\lambda t=m\pi (m=1,3,5,...)$. It
means that the requirement $\sqrt{1+\frac{\nu^2}{\lambda^2}}=n
(n=2,4,6,...)$ can be loosed in the limit $\nu \gg \lambda$. In
order to make the thing more clear, let us introduce the normal
modes
\begin{eqnarray}
\label{normal}
c_{0,j}&=&\frac{a_{A,j}-a_{B,j}}{\sqrt{2}},\\
c_{\pm,j}&=&\frac{a_{A,j}+a_{B,j}\pm \sqrt{2}b_j}{\sqrt{2}}.
\end{eqnarray}
In terms of these normal modes, the Hamiltonian (\ref{HI1}) can be
rewritten as
\begin{equation}
\begin{aligned}
\label{HI2} H_I=\sum_{j=\pm
1}\Big[&\lambda_{A,j}\frac{e^{i\sqrt{2}\nu_j
t}c_{-,j}+e^{-i\sqrt{2}\nu_j
t}c_{+,j}+\sqrt{2}c_{0,j}}{2}\sigma^{+}_{A,j}\\
&+\lambda_{B,j}\frac{e^{i\sqrt{2}\nu_j t}c_{-,j}+e^{-i\sqrt{2}\nu_j
t}c_{+,j}-\sqrt{2}c_{0,j}}{2}\sigma^{+}_{B,j}\Big].
\end{aligned}
\end{equation}
As noted in (16), the normal mode $c_{0,j}$ is resonant with the
atomic transitions from both $\vert e\rangle$ to $\vert g_j\rangle$
of the atom A and $\vert g\rangle$ to $\vert e_j\rangle$ of the atom
B, but the normal modes $c_{\pm,j}$ off-resonant. In the limit
$\nu_j\gg \lambda_{A,j},\lambda_{B,j}$,  the off-resonant modes can
be safely neglected. In this case,  the Hamiltonian (16) becomes
\begin{equation}
\label{HI3} H_I=\frac{1}{\sqrt{2}}\sum_{j=\pm
1}\Big[\lambda_{A,j}c_{0,j}\sigma^+_{A,j}
-\lambda_{B,j}c_{0,j}\sigma^+_{B,j}+\mathrm{H.c.}\Big].
\end{equation}
In the Hamiltonian (17), the fibre modes are completely depressed
since the resonant modes $c_0$ do not contain the fibre modes and
the original system is modeled by a system consisting of two atoms
interacting with two resonant modes in one cavity. For the initial
condition $\vert\psi_2(0)\rangle=\vert e\rangle_A\vert
g\rangle_B|0\rangle_c$, where $|0\rangle_c$ is the vacuum state of
the normal mode $c_{0}$, the solution of the Schr\"odinger
equation (7) with the Hamiltonian (17) can be found to be
\begin{equation}
\label{psi2}
\begin{aligned}
|\psi_2(t)\rangle&=\tilde{d_1}|e\rangle_A|0\rangle_c|g\rangle_B+
\tilde{d_2}|g_{-1}\rangle_A|1_{-1}\rangle_c|g\rangle_B+\tilde{d_3}|g_{+1}\rangle_A|1_{+1}\rangle_c|g\rangle_B\\
&+\tilde{d_4}|g_{-1}\rangle_A|0\rangle_c|e_{-1}\rangle_B+\tilde{d_5}|g_{+1}\rangle_A|0\rangle_c|e_{+1}\rangle_B,
\end{aligned}
\end{equation}
where $\vert 1_{\pm1}\rangle_c$ is a state in which there is
either one right(-1) or left(+1) circular polarized photon in the
normal mode $c_0$. Under the condition
$\lambda_{A,-1}=\lambda_{A,+1}=\lambda$, and
$\lambda_{B,-1}=\lambda_{B,+1}=\sqrt{2}\lambda$, the expansion
coefficients in (18) are given by
\begin{equation}
\label{coeff2}
\begin{aligned}
&\tilde{d_1}=\frac{1+cos(\sqrt{2}\lambda t)}{2},
\tilde{d_2}=\tilde{d_3}=\frac{-isin(\sqrt{2}\lambda t)}{2},
\tilde{d_4}=\tilde{d_5}=\frac{1-cos(\sqrt{2}\lambda
t)}{2\sqrt{2}}.
\end{aligned}
\end{equation}
From Eqs.(18) and (\ref{coeff2}), one can see that at the moments
$t=m\pi/\sqrt{2}\lambda(m=1,3,5,...)$ the system is in the state
$\vert\Psi_t\rangle\vert 0\rangle_c$ in which the two atoms are
separated from the normal modes and the extremely entangled state
of the atoms $\vert\Psi_t\rangle$ is achieved. This result depends
on the condition $\nu\gg\lambda$. In order to check the
approximation validity, in Fig.\ref{nodis}, we show the fidelities
$F_1=\vert\langle 0_A,0_F,0_B\vert\langle
\Psi_t\vert\psi_1(t)\rangle\vert^2$, where $\vert
0_A,0_F,0_B\rangle$ is the vacuum state for photons in both the
cavities and the fibre, and $F_2=\vert{}_c\langle
0\vert\langle\Psi_t\vert\psi_2(t)\rangle\vert^2$ that corresponds
to the limit $\nu\rightarrow\infty$.
As shown in Fig.\ref{nodis}, the results related to the Hamiltonians
(6) and (17) become nearly same and the approximation condition
$\nu\gg\lambda$ holds very well when $\nu\geqq10\lambda$.

From the results above, we come to the conclusion that to generate
the extremely entangled state of the atoms  it is needed to choose
the right interaction time and the right ratio of $\nu$ to $\lambda$
when $\nu$ is comparable to $\lambda$, but it is needed only to
control the interaction time when $\nu\gg\lambda$.

\section{Effects of spontaneous emission and photon loss}

In this section, we investigate the influence of spontaneous
emission and photon leakage on the generation of atomic entangled
states. The master equation for the density matrix of the whole
system is
\begin{equation}
\begin{aligned}
\dot{\rho}=-i[H_I,\rho]
&-\sum_{j=-1,1}\frac{\gamma_{A,j}}{2}(a^+_{A,j}a_{A,j}\rho-2a_{A,j}\rho
a^+_{A,j}+\rho a^+_{A,j}a_{A,j})\\
&-\sum_{j=-1,1}\frac{\gamma_{B,j}}{2}(a^+_{B,j}a_{B,j}\rho-2a_{B,j}\rho
a^+_{B,j}+\rho a^+_{B,j}a_{B,j})\\
&-\sum_{j=-1,1}\frac{\gamma_{f,j}}{2}(b^+_{j}b_{j}\rho-2b_{j}\rho
b^+_{j}+\rho b^+_{j}b_{j})\\
&-\sum_{j=-1,1}\frac{\kappa_{A,j}}{2}(\sigma^+_{A,j}\sigma_{A,j}\rho-2\sigma_{A,j}\rho
\sigma^+_{A,j}+\rho \sigma^+_{A,j}\sigma_{A,j})\\
&-\sum_{j=-1,1}\frac{\kappa_{B,j}}{2}(\sigma^+_{B,j}\sigma_{B,j}\rho-2\sigma_{B,j}\rho
\sigma^+_{B,j}+\rho \sigma^+_{B,j}\sigma_{B,j}),
\end{aligned}
\end{equation}
where $H_I$ is given by (6), and $\gamma_{A,j}, \gamma_{B,j}$ and
$\gamma_{f,j}$ are the decay rates for photons in mode $j$ of
cavities A and B, and fibre, respectively, and $\kappa_{A,j}$ and
$\kappa_{B,j}$ are spontaneous emission rates that are related to
the decay channel of atom A: $\vert e\rangle\rightarrow\vert
g_j\rangle$ and the decay channel of atom B: $\vert
e_j\rangle\rightarrow\vert g\rangle$, respectively. In solving the
master equation, we always choose $\lambda_{B,\pm
1}=\sqrt{2}\lambda_{A,\pm 1}=\sqrt{2}\lambda$, and $\kappa_{A,\pm
1}=\kappa_{B,\pm 1}=\kappa_a$, $\gamma_{f,\pm 1}=\gamma_f$, and
$\gamma_{A,\pm 1}=\gamma_{B,\pm 1}=\gamma_c$. By use of the solution
of Eq. (20), the fidelity $F_3=\langle
0_A,0_F,0_B\vert\langle\Psi_t\vert \rho(t)\vert\Psi_t\rangle\vert
0_A,0_F,0_B\rangle$ can be calculated.
\begin{figure}[htbp]
 \centering
\mbox{\subfigure[]{\includegraphics[width=7.5cm]{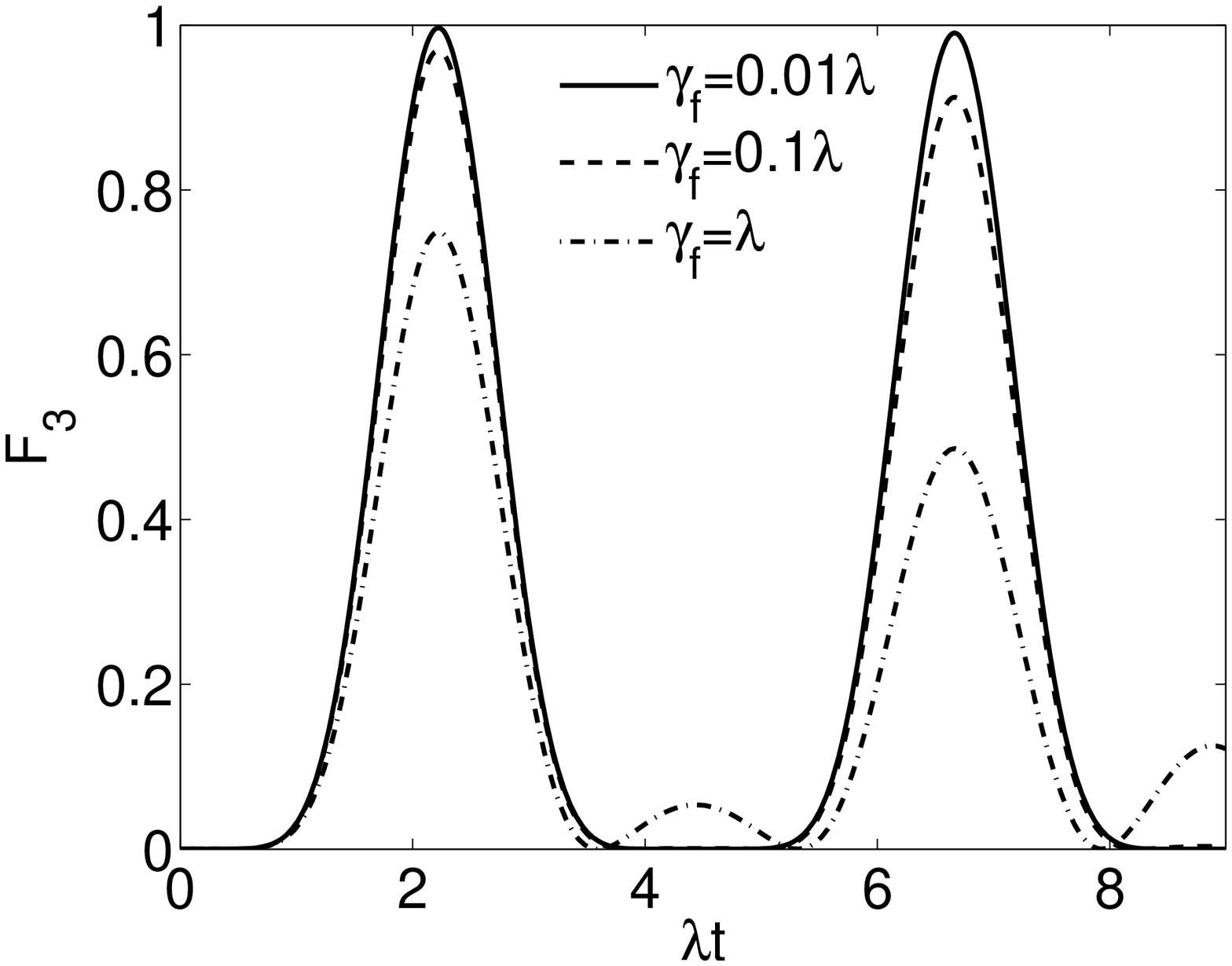}}
\subfigure[]{\includegraphics[width=7.5cm]{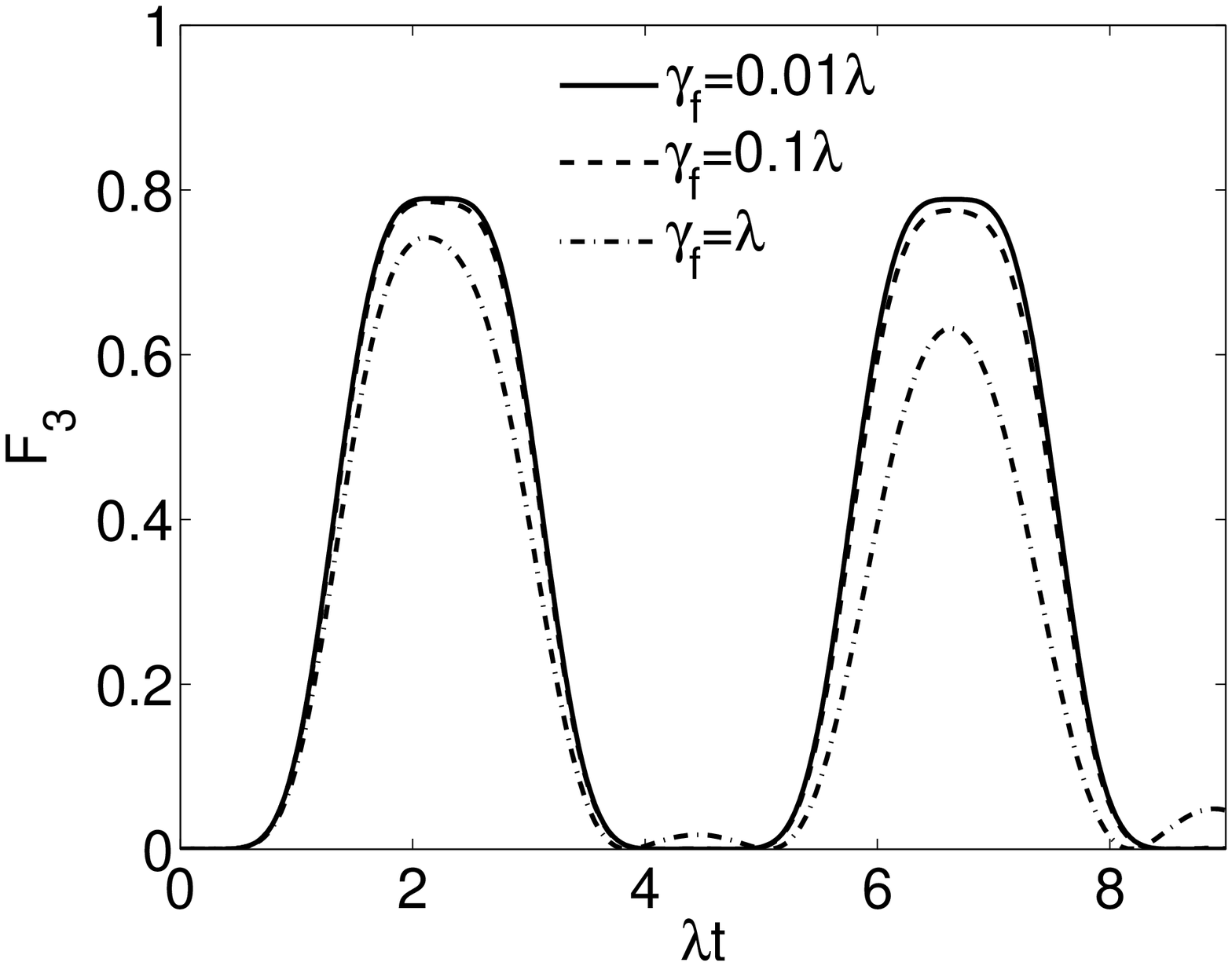}}}\\
\mbox{\subfigure[]{\includegraphics[width=7.5cm]{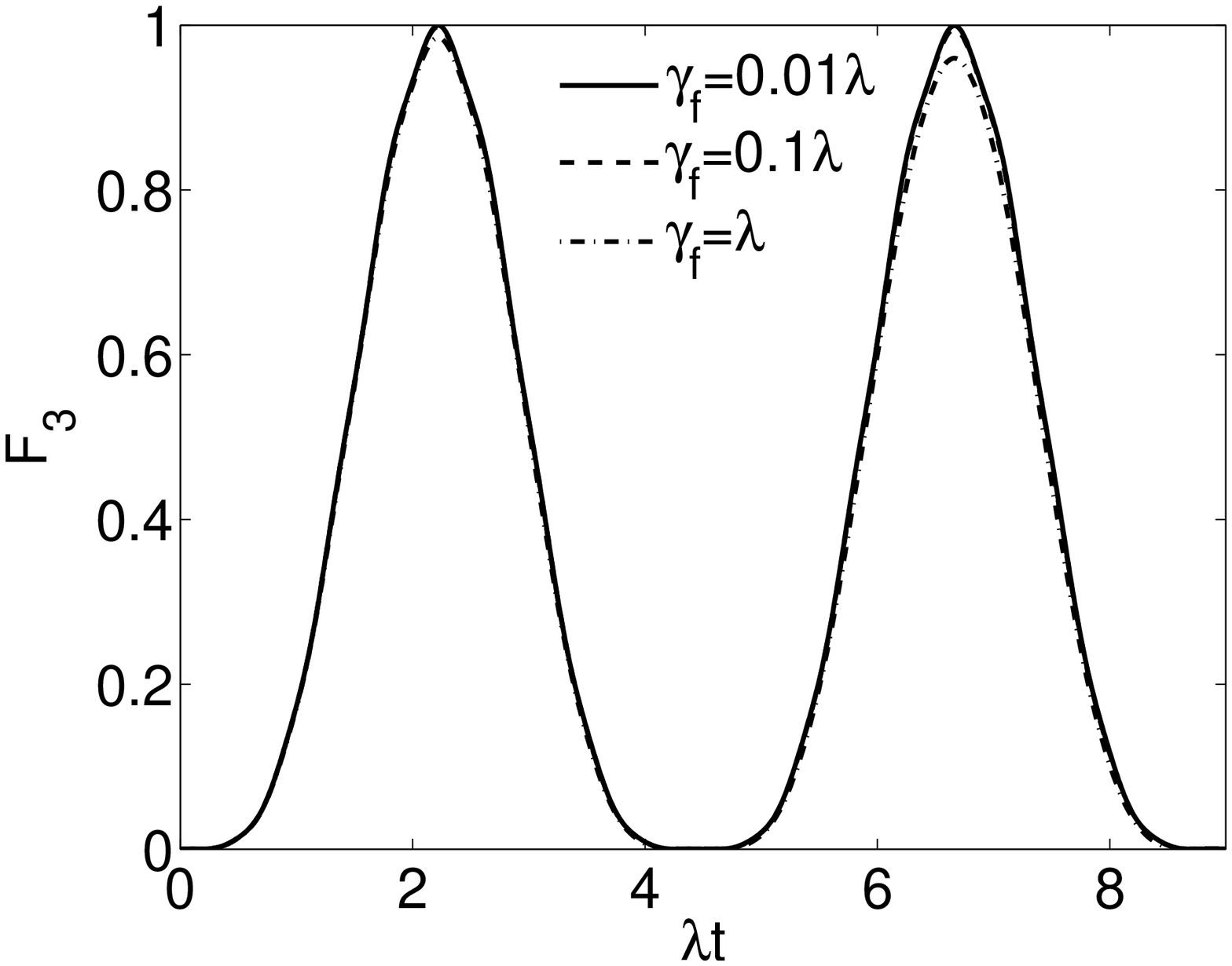}}
\subfigure[]{\includegraphics[width=7.5cm]{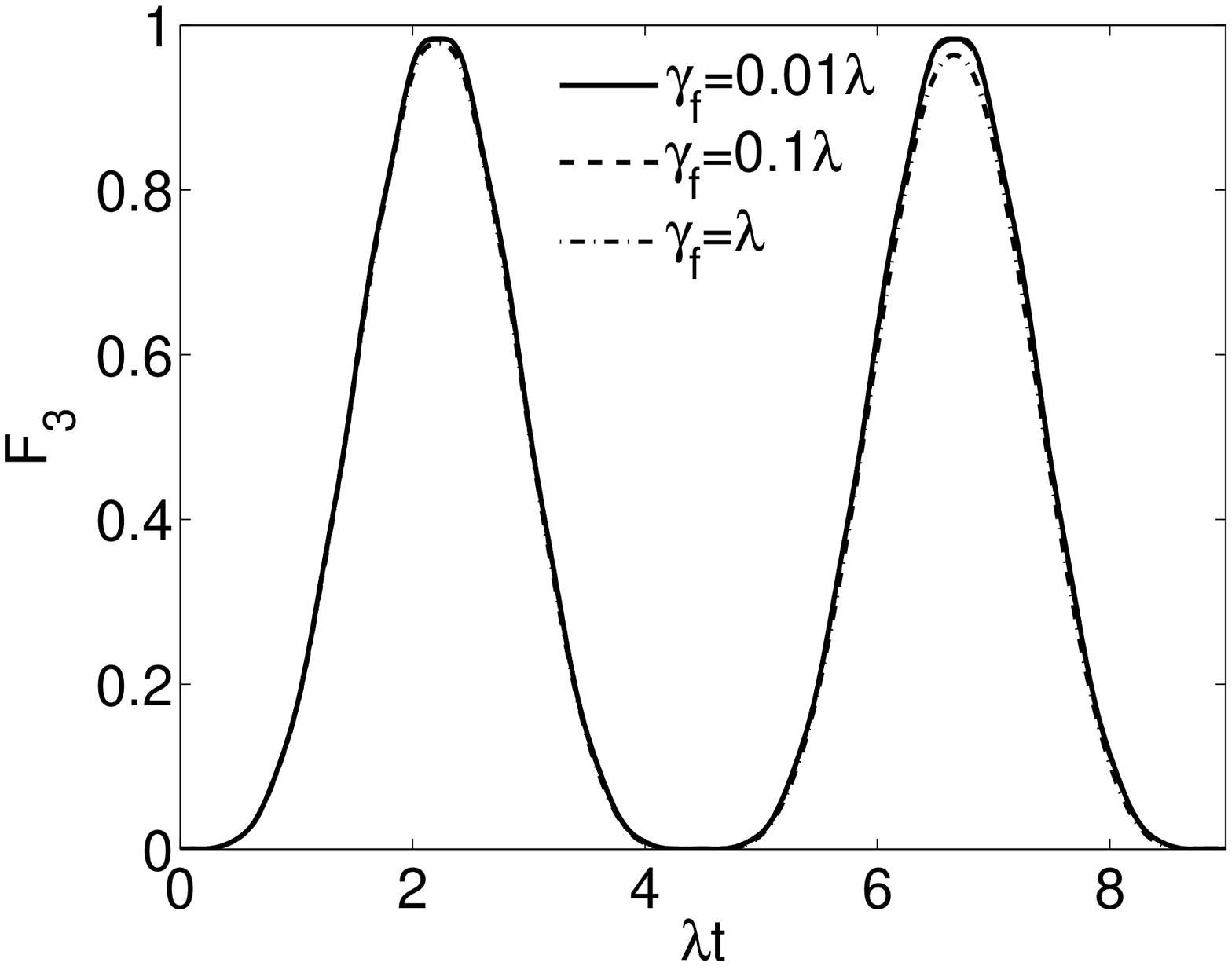}}}
       \caption{(a)$\nu=\sqrt{3}\lambda$;(b)$\nu=2\sqrt{2}\lambda$;(c)$\nu=\sqrt{99}\lambda$;(d)$\nu=\sqrt{120}\lambda$}
       \label{rfdis}
\end{figure}
\begin{figure}[htbp]
 \centering
\mbox{\subfigure[]{\includegraphics[width=7.5cm]{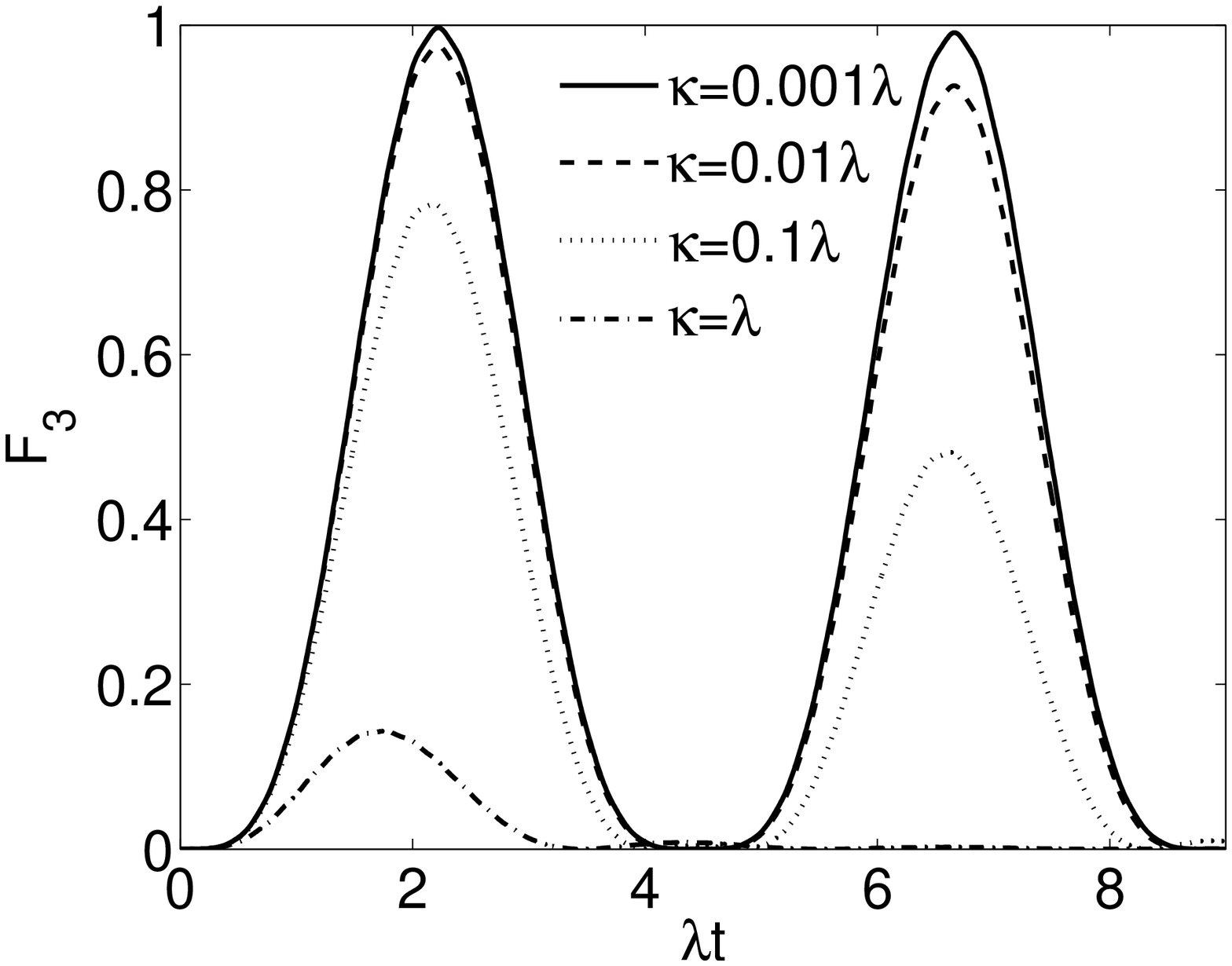}}
\subfigure[]{\includegraphics[width=7.5cm]{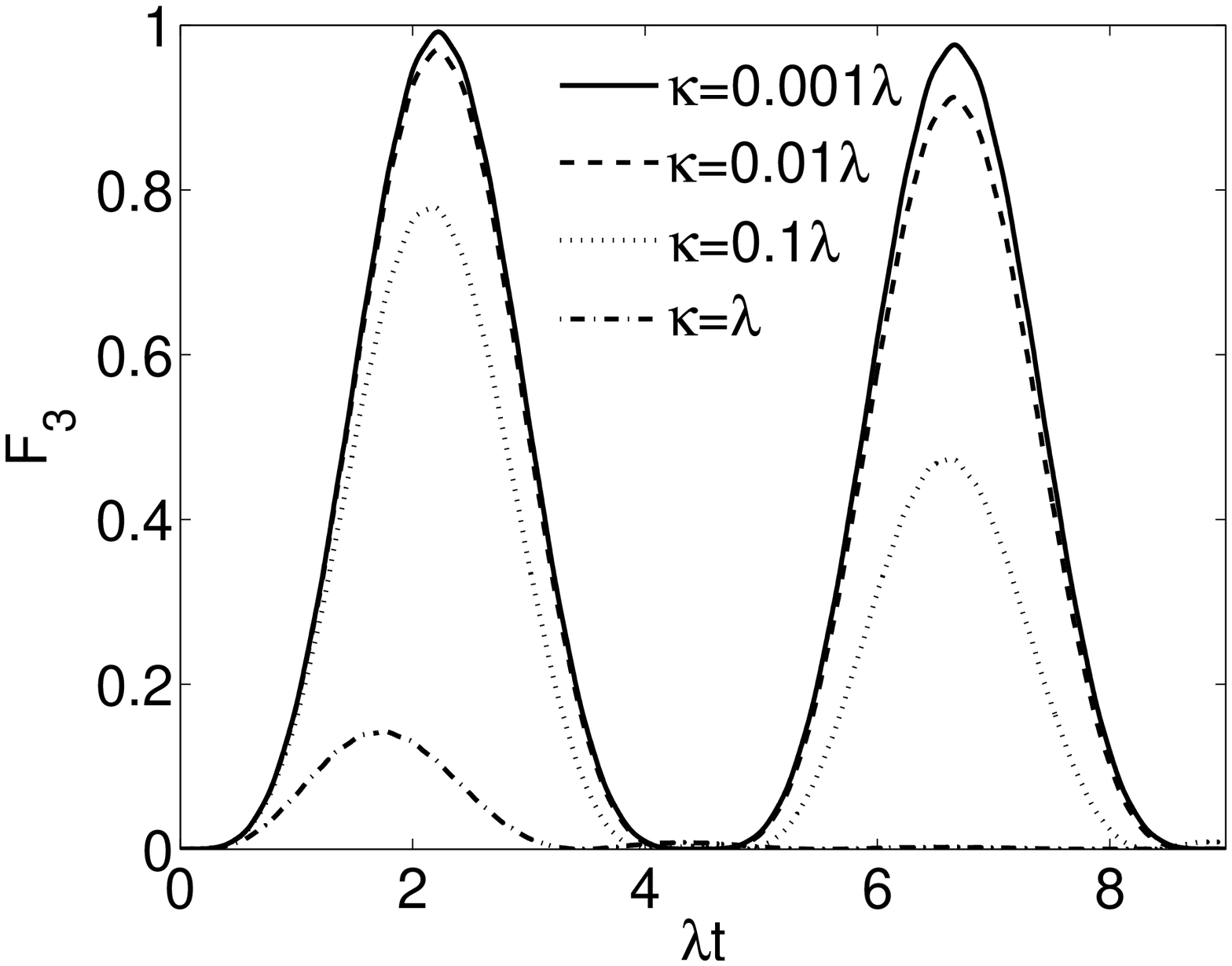}}}\\
\mbox{\subfigure[]{\includegraphics[width=7.5cm]{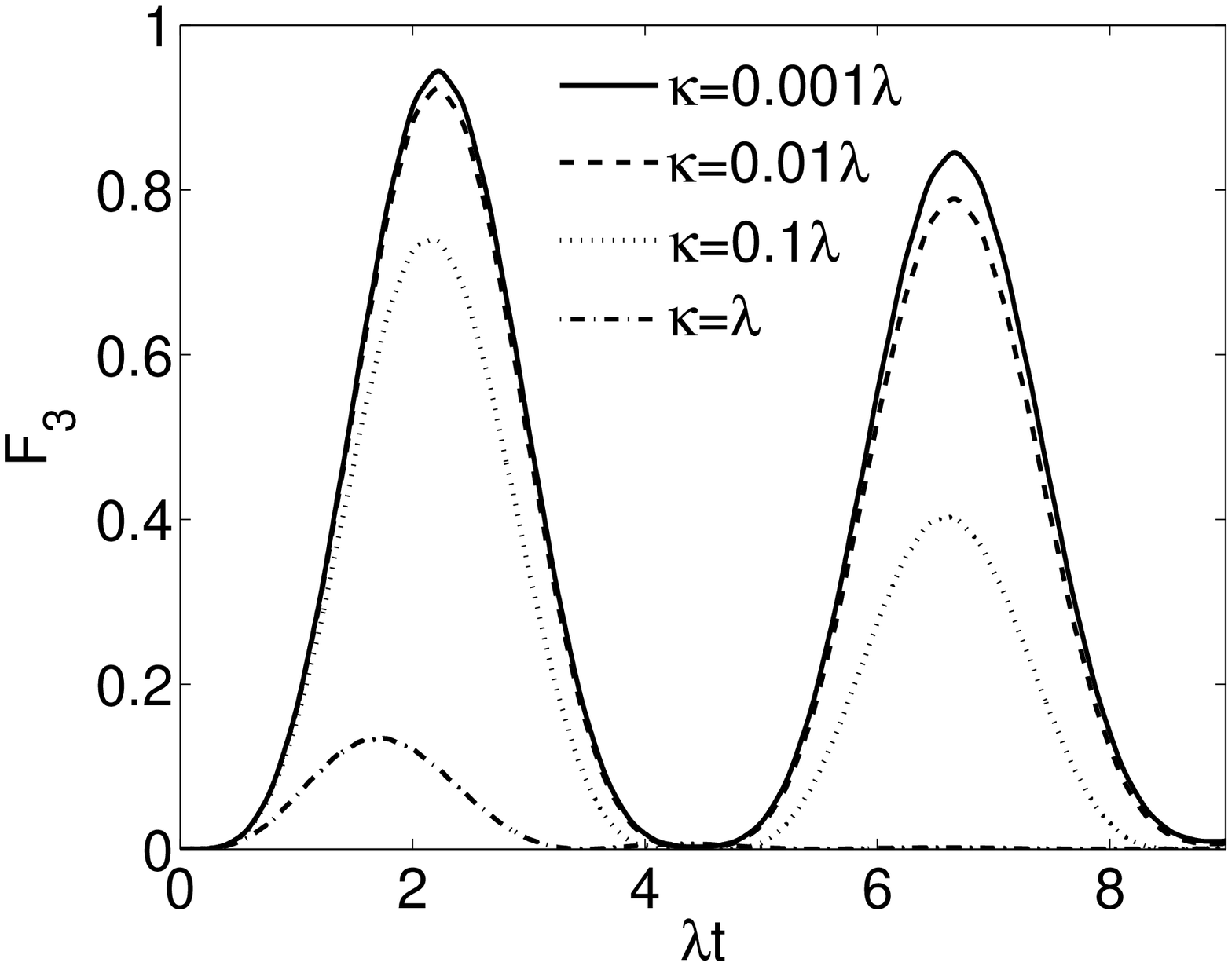}}
\subfigure[]{\includegraphics[width=7.5cm]{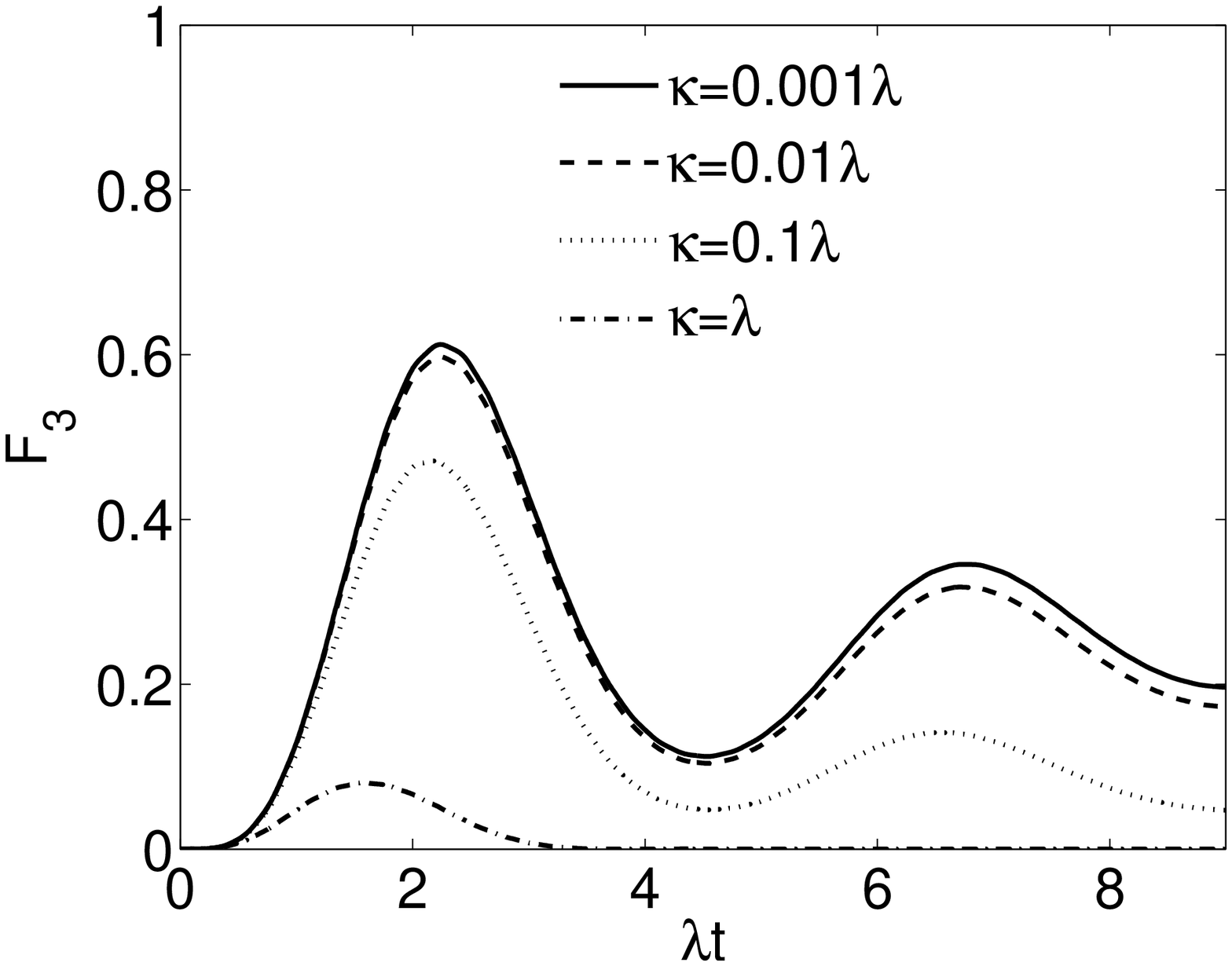}}}
       \caption{(a)$\gamma=0.001\lambda$(b)$\gamma=0.01\lambda$;(c)$\gamma=0.1\lambda$;(d)$\gamma=\lambda$}
       \label{rkdis}
\end{figure}
In Fig.\ref{rfdis}, the fidelity $F_3$ is plotted for the cases with
$\kappa_a=\gamma_c=0$ but different values of $\nu$ and $\gamma_f$.

As shown in Fig. 3(c) and (d), even for the case with
$\gamma_f=\lambda$, the effect of photon leakage out of the fibre
can be greatly depressed when the cavity-fiber coupling strength
$\nu$ is much larger than the cavity-atom coupling strength
$\lambda$. This result can be easily understood. From the discussion
in the preceding section, there is only the resonant normal mode
$c_0$ plays role when $\nu\gg\lambda$. According to (14), the
resonant normal mode is not involved with the fibre mode. In the
limit, therefore, the fibre mode is not excited and is always kept
in the vacuum state. On the other hand, the condition
$\nu\gg\lambda$ means that the transmission time of photons through
the fibre becomes much shorter than the interaction time of the
atoms with the cavity fields and then the fibre photon is
equivalently kept in the vacuum state in the whole process. This
reason clearly explains why the influence of photon loss can be
diminished in the limit $\nu\gg\lambda$.

From Eqs. (9)-(13), $d_3=d_7=0$ at
$t=m\pi/\sqrt2\lambda(m=1,3,5,...)$ if the condition
$\sqrt{1+\frac{\nu^2}{\lambda^2}}=n (n=2,4,6,...)$ is satisfied. It
means that all the terms associated with the basic vectors with one
fibre photon in (8) disappear under this condition. In this case,
therefore, the fibre field in the state vector (8) is in the vacuum
state. Therefore, the appropriate choice of the ratio $\nu/\lambda$
may be useful to get a higher fidelity of the entangled state even
if the condition $\nu\gg\lambda$ is not well satisfied. In Fig. 3(a)
and (b), the results with $\sqrt{1+\frac{\nu^2}{\lambda^2}}=2$ and
$\sqrt{1+\frac{\nu^2}{\lambda^2}}=3$ are shown. It is observed that
the former result is obviously better than the latter one when
$\gamma_f < 0.1\lambda$.

Next, we investigate the influence of spontaneous emission and
cavity leakage. In Figs.\ref{rkdis}, the fidelity $F_3$ is shown
for the cases with $\nu=\sqrt{399}\lambda$ and $\gamma_f=0$, and
different values of $\gamma_c$ and $\kappa_a$. Comparing these
figures, we find that the atomic spontaneous emission has a
stronger influence on the fidelity than the cavity photon leakage
does. To obtain the fidelity higher than 0.97, one should keep the
decay rates $\kappa_a\leq 0.01\lambda$ and $\gamma_c\leq
0.1\lambda$. In recent experiments \cite{Spillane,buck}, for an
optical cavity with the wavelength region $630nm\sim850nm$, the
condition $(\lambda/2\pi=750MHz, \kappa_a/2\pi=2.62MHz,
\gamma_c/2\pi=3.5MHz)$ is achievable. Therefore, the present
scheme with a high fidelity larger than 0.98 is feasible in
current experiments.

\section{conclusions}

In this paper, we propose a scheme to generate an entangled state of
two atoms trapped in spatially separated two cavities that are
connected by an optical fibre. We show that an extremely entangled
state of the atoms can be deterministically generated if the ratio
of the coupling constant of the cavity field with the atoms to the
coupling constant of the cavity field with the fibre mode satisfies
a certain condition. The scheme has three features. First, it is
based on both the photon emission of a $\Lambda$-type atom and the
photon absorption of a V-type atom. Therefore, in contrast to the
previous schemes \cite{Lloyd,Feng,Duan}, either a pair of entangled
photons or the photon interference process is not required. Second,
since the present scheme is based on the interaction dynamics of
atoms with cavity fields and then the co-instantaneous single photon
detection is not required, the generation of entangled states is
deterministic. Third, in contrast to the previous schemes in which
manipulations such as excitation, photon emission or absorption have
to be performed simultaneously on all of atoms under consideration,
such manipulations are performed respectively on each of atoms in
the present scheme. This may reduce difficulties resulting from
manipulations performed at the same time on two atoms. The influence
of various decoherence processes such as spontaneous emission and
photon loss on the generation of entangled states is also
investigated. We find that the effect of photon leakage out of the
fibre can be greatly depressed and even ignored if the coupling of
the fibre mode with the cavity fields is much stronger than the
coupling of the atoms with the cavity fields. If the condition is
not well satisfied, the appropriate ratio of the coupling constant
of the cavity field with the atoms to the coupling constant of the
cavity field with the fibre mode can also ensure a high fidelity of
generating entangled states. As regards the effect of photon loss
out of the cavities and spontaneous emission of the atoms, we find
that the implementation of the present scheme is within the scope of
the current techniques.

\acknowledgments  The authors thank Zhang-qi Yin and Yang Yang for
many useful discussions. This work was supported by the Natural
Science Foundation of China (Grant Nos. 10674106, 10574103,
05-06-01).

\end{document}